\documentclass[aps,prb,superscriptaddress,twocolumn,showpacs,floatfix]{revtex4}

\usepackage{amsmath}
\usepackage{amssymb}
\usepackage{bm, color}
\usepackage{graphicx,epsfig,color}

\begin{document}

\title{Optical bistability and hysteresis of hybrid metal-semiconductor nano-dimer}

\author{A.\ V.\ Malyshev}
\email{a.malyshev@fis.ucm.es}

\affiliation{GISC, Departamento de F\'{\i}sica de Materiales,
Universidad Complutense, E-28040 Madrid, Spain} \affiliation{Ioffe
Physical-Technical Institute, 26 Politechnicheskaya str., 194021
St.-Petersburg, Russia}

\affiliation{Centre for Theoretical Physics and Zernike Institute
for Advanced Materials, University of Groningen, Nijenborgh 4, 9747
AG Groningen, The Netherlands}

\author{V.\ A.\ Malyshev}
\affiliation{Centre for Theoretical Physics and Zernike Institute
for Advanced Materials, University of Groningen, Nijenborgh 4, 9747
AG Groningen, The Netherlands}

\affiliation{St. Petersburg State University, 198504 St. Petersburg,
Russia}

\date{\today}

\begin{abstract}

Optical response of an artificial composite nano-dimer comprising a
semiconductor quantum dot and a metal nanosphere is analyzed theoretically. 
We show that internal degrees of freedom of the system can manifest
bistability and optical hysteresis as functions of the incident field
intensity.  We argue that these effects can be observed for the real world
systems, such as a CdSe quantum dot and an Au nanoparticle hybrids.  These
properties can be revealed by measuring the optical hysteresis of the
Rayleigh scattering.  We show also that the total dipole moment of the
system can be switched abruptly between its two stable states by small
changes in the excitation intensity.  The latter promises various
applications in the field of all-optical processing at nanoscale, the most
underlying of them being the volatile optical memory.

\end{abstract}


\maketitle

\section{Introduction}

Arrays of metallic nanoparticles (often referred to as plasmonic arrays),
are widely recognized as potential building blocks for nanoscale optical
circuitry~\cite{Quinten98, Brongersma00, Li03, Stockman04, Citrin04,
Hernandez05, Waele07, Malyshev08, Zheng09} (see also
Ref.~\onlinecite{Maier07} for an overview).  Recently, a number of groups
reported fascinating properties of artificial molecules comprised of a
semiconductor quantum dot (SQD) in the proximity of a metallic nanoparticles
(MNP).~\cite{Pons07,Zhang06,Govorov06,Artuso08,Artuso10,Sadeghi09a,
Sadeghi09b,Sadeghi10a,Sadeghi10b} Nonlinear Fano
resonances~\cite{Zhang06,Artuso08} and bistability in the absorption
spectrum,~\cite{Artuso08,Artuso10} control of the exciton emission of the
SQD (inhibition or enhancement)~\cite{Govorov06} and variable quenching of
the SQD photoluminescence by proximate gold nanoparticles,~\cite{Pons07} as
well as ``meta-molecular`` resonances,~\cite{Sadeghi09a} the inhibition of
optical excitation and enhancement of Rabi flopping,~\cite{Sadeghi09b}
tunable nanoswitching,~\cite{Sadeghi10a} and gain without
inversion~\cite{Sadeghi10b} have been predicted.  The role of the multipole
SQD-MNP interaction in explaining the spectra of hybrid systems has been
discussed in details in Ref.~\onlinecite{Govorov08}.  All these effects
depend on both geometrical parameters and material properties of hybrid
clusters providing an excellent opportunity of more fine-grained control of
spectral and dynamical properties of nanoscale objects.

We consider the simplest hybrid nano-cluster comprising a SQD and a
spherical MNP -- artificial hybrid diatomic nano-molecule.  When such a
system is excited optically, the dipole moment of the optical transition in
the SQD generates an additional electric field at the MNP, which is
superposed with the external field.  Similarly, the induced dipole moment of
the MNP generates an additional electric field in the SQD.  Thus, the
presence of the MNP leads to a self action (feedback) of the SQD.  Together
with nonlinearity of the SQD itself, this can give rise to a variety of new
optical properties.  In particular, if the coupling between two
nanoparticles is strong enough, the self-action can result in optical
bistability of the response.  Note that a dimer comprised of strongly
coupled two-level molecules can not manifest bistability.~\cite{Malyshev98}
Thus, a SQD-MNP hetero-dimer is a fascinating nanoscopic system exhibiting
this feature.

To demonstrate the feasibility of the bistable optical response of hybrid
composites, we consider a closely spaced CdSe (or CdSe/ZnSe) SQD and an Au
nano-sphere.  We show that for a range of geometrical parameters of the
system (SQD and MNP radii and center-to-center distance), the optical
bistability and hysteresis can be observed in it.  We argue also that
because of the axial symmetry of such artificial diatomic molecule, its
state can be switched not only by traditional change of the driving field
amplitude, but also by changing the incoming field polarization with respect
to the molecule axis, which offers an additional mechanism of control.  The
fact that both the SQD and the MNP can sustain high electric fields suggests
such possible applications of artificial molecules as all-optical switches
and optical memory cells at nanoscale in the visible; the two stable states
of the systems have different total dipole moments, providing a possibility
to store information in this degree of freedom.

The paper is organized as follows. In the next Section the model and
formalism are described.  In Sec.~\ref{Steady-state analysis} we
present the standard steady-state analysis of the bistable optical
response, which gives rise to the optical hysteresis addresses in
the Section~\ref{Optical hysteresis}. We discuss possible
applications of the predicted effects in Sec.~\ref{Rayleigh
scattering and optical storage} while Section~\ref{Summary}
summarizes the paper.

\begin{figure}[ht]
\begin{center}
         \includegraphics[width=0.45\columnwidth]{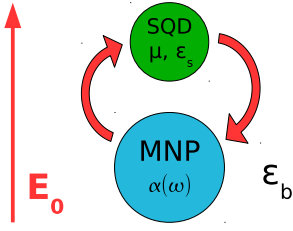}
\end{center}
\caption{Schematics of the hybrid SQD-MNP system embedded into a
homogeneous dielectric background with the permittivity $\varepsilon_b$ and
subjected to an external field with an amplitude ${\bf E}_0$.  The SQD
optical transition dipole moment and the semiconductor dielectric constant
are denoted as ${\bm \mu}$ and $\varepsilon_s$, respectively, while $\alpha
(\omega)$ is the MNP polarizability.  The curved arrows symbolize the
dipole-dipole SQD-MNP interaction.}
\label{Schematics}
\end{figure}

\section{Formalism}
\label{Formalism}

We assume that the SQD-MNP hybrid molecule is embedded in a dielectric host
with the permittivity $\varepsilon_b$ and is driven by a linearly polarized
external electric field with an amplitude ${\bf E}_0$ and frequency
$\omega$.  Figure~\ref{Schematics} shows the schematics of the system.  The
SQD is modeled as a two level system with the transition frequency
$\omega_0$ and optical transition dipole moment $\bm\mu$.  It is treated
quantum mechanically within the framework of the Maxwell-Bloch equations for
the $2\times 2$ density matrix $\rho_{mn} \> (m,n = 0,1)$.  The MNP is
considered classically; the response of the MNP is described by its
frequency dependent scalar polarizability within the point dipole
approximation (this can easily be generalized for the case of more complex
shapes of the MNP by considering an appropriate polarizability tensor).  All
sizes of the system (the SQD and MNP radii and the SQD-MNP center-to-center
distance) are assumed to be small enough to neglect the retardation effects
and to consider both particles as point dipoles.  The rotating wave
approximation is used throughout the paper, so that all time dependent
quantities represent amplitudes of the corresponding characteristics of SQD,
the set of equations for which reads:
\begin{subequations}
\begin{equation}
\label{dotZ} \dot Z = - \gamma\,(Z+1)-\frac{1}{2} \left[\,
\Omega\,R^{*}+\Omega^{*}R \,\right] \ ,
\end{equation}
\begin{equation}
\label{dotR} \dot R = -(i\,\Delta+\Gamma)\, R+\Omega\,Z \ ,
\end{equation}
\end{subequations}
where $Z=\rho_{11}-\rho_{00}$ is the population difference between
the SQD excited and ground states, $R$ is the amplitude of the
off-diagonal density matrix element defined through $\rho_{10} =
-(i/2)R \exp(-i\omega t)$, $\gamma$ and $\Gamma$ are the relaxation constant
of the population and the dipole dephasing, respectively, $\Delta = \omega_0
- \omega$ is the detuning of the driving field from the SQD resonance, and
$\Omega={\bm\mu\bf E}/\hbar$ is the electric field (in frequency units)
acting {\it inside} the SQD.  The field acting upon the SQD is equal to the
sum of the external field ${\bf E}_0$ and the field produced by the induced
dipole moment ${\bf P}_\mathrm{MNP}$ of the MNP.  The field ${\bf E}$ {\it
inside} the SQD is reduced by the factor $\varepsilon_s^{\prime} =
(\varepsilon_s + 2\varepsilon_b)/(3\varepsilon_b)$ where $\varepsilon_s$ is
the permeability of the SQD (see e.  g.  Ref.~\onlinecite{Batygin70} or
Ref.~\onlinecite{Bohren83}, Ch.  V, p.  138):
\begin{equation}
\label{E_SQD}
{\bf E}=\frac{1}{\varepsilon_{\mathrm{s}}^{\prime}}
\left( {\bf E}_0+\frac{ {\widehat{\bf S}}\,{\bf P}_{\mathrm{MNP}}
}{\varepsilon_b\,d^3} \right) \ .
\end{equation}
Here, ${\widehat{\bf S}} = \mathrm{diag}(-1,-1,2)$ is the angular
part of the dipole field Green's tensor ($z$ axis being parallel to
the system axis), $d$ is the SQD-MNP center-to-center distance while
${\bf P}_\mathrm{MNP}$ is given by:
\begin{equation}
\label{P_MNP}
{\bf P}_{\mathrm{MNP}}=\varepsilon_b\,\alpha(\omega)
\left(
{\bf E}_0+\frac{{\widehat{\bf S}}\,{\bf P}_{\mathrm{SQD}}}{\varepsilon_b\,d^3}
\right) \ ,
\end{equation}
where $\alpha(\omega)=a^3\,\gamma(\omega)$ is the classical frequency
dependent polarizability of the MNP, $a$ being its radius, $\gamma(\omega)=
[\varepsilon_M(\omega)-\varepsilon_b]/[\varepsilon_M(\omega)+
2\varepsilon_b]$, and $\varepsilon_M(\omega)$ is the dielectric function of
the metal.  We do not take into account the corrections to the
polarizability due to the depolarization shift and radiative
damping,~\cite{Meier83} which are negligible for nanoparticle sizes of our
interest ($\leq$10 nm).  The second term in the parenthesis of
Eq.~(\ref{P_MNP}) is the field produced by the SQD dipole moment ${\bf
P}_{\mathrm{SQD}}=-i{\bm\mu}\,R$ at the MNP.  

The exciton radius in CdSe is about 5 nm~\cite{Ekimov93} while the typical
radius of the considered SQD is about $1.5-2$ nm, so the wavefunctions
involved in the optical transition are extended over the whole dot.  In
deriving Eq.~(\ref{P_MNP}) we used therefore the approximation of the
homogeneous electric polarization of the whole SQD volume.  In this case the
dipole field around the SQD is screened by the bare background dielectric
constant only.~\cite{Batygin70}  Note that the dipole moment ${\bf
P}_{\mathrm{SQD}}$ is calculated quantum mechanically and accounts for the
screening which results from the SQD dielectric response (see below). 
Finally, for the total electric field {\it inside} the SQD we obtain:
\begin{equation}
\label{E SQD 1} {\bf E}= \frac{1}{\varepsilon_{\mathrm{s}}^{\prime}}
\left[ {\bf 1}+\frac{\gamma(\omega)\,a^3}{d^3}\,{\widehat{\bf S}}
\right]{{\bf E}_0} +
\frac{\gamma(\omega)\,a^3}{\varepsilon_b\,\varepsilon_s^{\prime}\,
d^6}\,{\widehat{\bf S}}^2\,{\bf P}_{\mathrm{SQD}}\ .
\end{equation}
As is seen from Eq.~(\ref{E SQD 1}), the presence of the MNP results in two
effects: the first term accounts for the renormalization of the external
field amplitude ${\bf E}_0$ while the second represents the self-action of
the SQD via the MNP; the field inside the SQD depends on the dipole moment
of the SQD itself.

The effect of the self-action on the dynamics of the SQD-MNP hybrid
nano-molecule can be revealed after substituting Eq.~(\ref{E SQD 1})
into Eq.~(\ref{dotR}) and representing $\Omega $ in the form
\begin{equation}
\label{Omega}
\Omega = \widetilde\Omega_0 - i\,G\,R
\end{equation}  
with $\widetilde\Omega_0$ and $G$ given by
\begin{subequations}
\begin{equation}
\label{Omega0prime}
\widetilde\Omega_0 = \frac{1}{\varepsilon_s^{\prime}} \left[ 1 +
\frac{a^3\gamma(\omega)}{d^3}\, \frac{\bm\mu\,\widehat{\bf S}\,{\bf
E}_0}{\hbar\Omega_0} \right] \Omega_0 \ ,
\end{equation}
\begin{equation}
 \label{G}
G = \frac{\gamma(\omega)\,a^3}
{\varepsilon_b\,\varepsilon_s^{\prime}\,\hbar\,d^6}\,{\bm\mu\,\widehat{\bf S}}^2\bm\mu \ ,
\end{equation}
\end{subequations}
where $\Omega_0 = {\bm\mu}{\bf E}_0/\hbar$ is the Rabi frequency of the bare
external field, $\widetilde\Omega_0$ is the renormalized Rabi frequency and
$G$ is the feedback parameter.  The latter absorbs all information governing
the SQD self-action, such as, material constants, geometry of the system or
details of the interaction ({\it e.  g.} contributions of higher
multipoles~\cite{Govorov08}).  

In a number of recent publications dealing with the same system, a different
formula for the constant $G$ was used in which the factor
$\varepsilon_{\mathrm{s}}^{\prime}$ appears squared in the denominator of
the $G$.~\cite{Zhang06,Artuso08,Artuso10,Govorov06,
Sadeghi09a,Sadeghi09b,Sadeghi10a,Sadeghi10b,Govorov08} The second factor
$\varepsilon_{\mathrm{s}}^{\prime}$ is supposed to take into account the
screening of the dipole field by the SQD dielectric response.  We note,
however, that it is the product $G\,R$ that determines the latter field and,
as can be seen from Eq.~(\ref{steady state R}), $R \propto
\widetilde\Omega_0 \propto \Omega_0/\varepsilon_{\mathrm{s}}^{\prime}$, so
the dipole field $G\,R$ is already additionally screened.

The feedback $G$ is the most important parameter of the theory; once it is
calculated it determines the nonlinear properties of the SQD response. 
Using Eq.~(\ref{Omega}), Eq.~(\ref{dotR}) can be rewritten in the following
form:
\begin{equation}
 \label{RR}
\dot R = -[(\Gamma - G_\mathrm{I}\,Z)+i\,(\Delta + G_\mathrm{R}\,Z)]\, R
+ \widetilde\Omega_0\,Z
\end{equation}
with $G_\mathrm{R} = \mathrm{Re} (G)$ and $G_\mathrm{I} = \mathrm{Im} (G)$. 
From Eq.~(\ref{RR}) two consequences of the SQD self-action become apparent:
(i) - the renormalization of the SQD resonance frequency $\omega_0 \mapsto
\omega_0 + G_\mathrm{R}\,Z$ and (ii) - the renormalization of the dipole
dephasing rate $\Gamma\mapsto\Gamma - G_\mathrm{I}\,Z$.  Both renormalized
quantities depend on the population difference $Z$.  Similar
renormalizations originate from the local field correction in the nonlinear
optical response of dense gaseous assemblies of two-level
systems,~\cite{Friedberg89} optically dense thin films,~\cite{Benedict90}
and linear molecular aggregates.~\cite{Malyshev96} The population
dependencies of the SQD resonance frequency and the dipole dephasing rate
provide a feedback mechanism, which results in a number of fascinating
effects.

\begin{figure}[ht]
\begin{center}
\includegraphics[width=0.35\columnwidth]{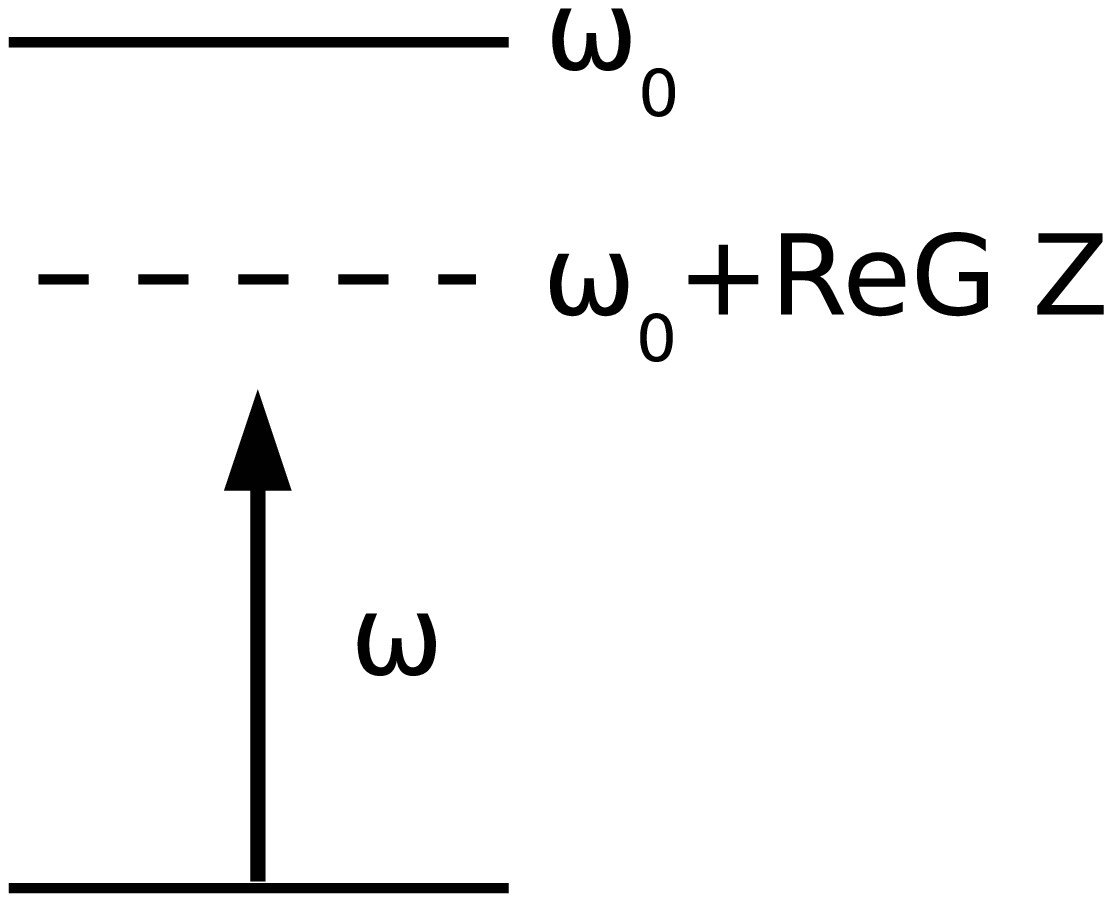}
\hskip 10pt
\includegraphics[width=0.35\columnwidth]{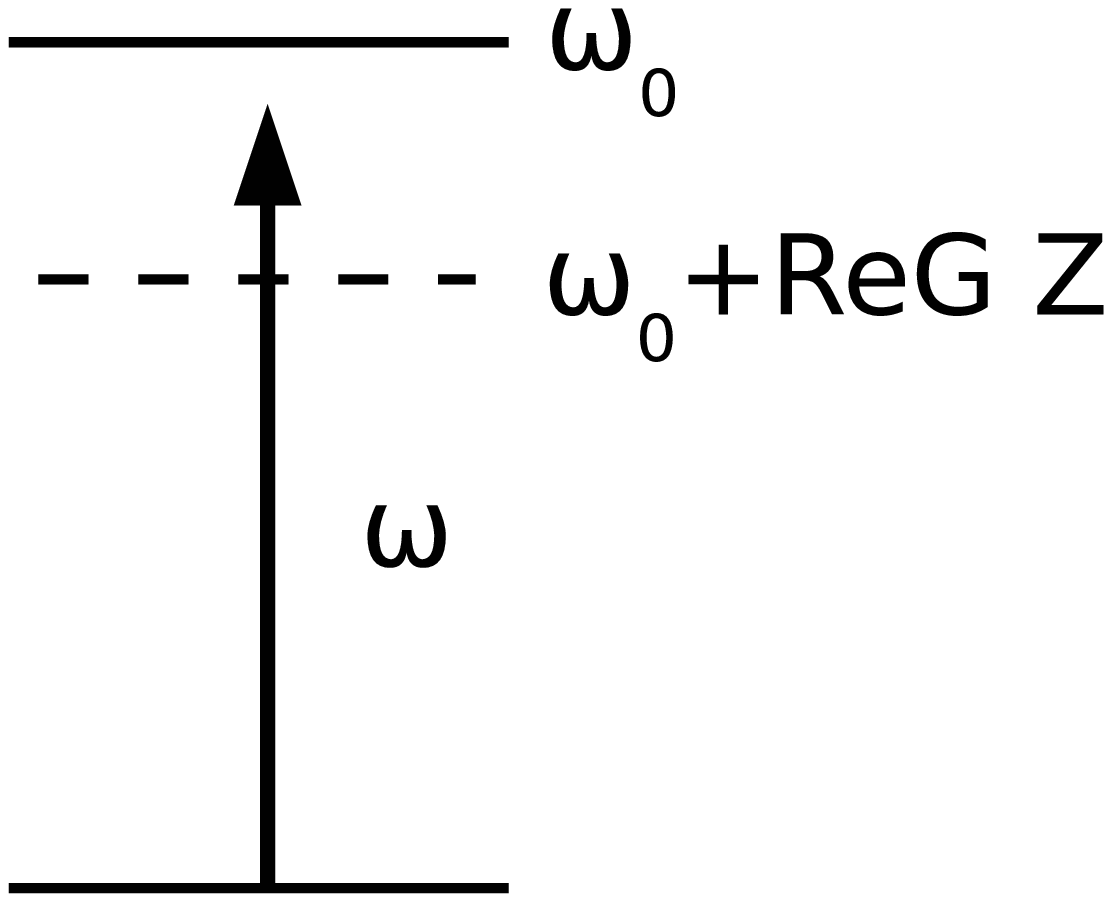}
\end{center}
    \caption{Two possible types of the SQD excitation. Left plot --
    the excitation frequency $\omega$ lies below the renormalized
    SQD transition frequency $\omega_0 - G_\mathrm{R}$ ($\Delta >
    G_\mathrm{R}$); as $Z$ increases with the excitation the system is
    driven further out of resonance. Right plot: $\omega > \omega_0
    - G_\mathrm{R}$ ($\Delta < G_\mathrm{R}$); the excitation drives the
    system into the resonance which favors the bistability to occur.
}
\label{Fig2}
\end{figure}

Let us assume that both $G_\mathrm{R} > 0$ and $G_\mathrm{I} > 0$, which is
the case for a CdSe SQD conjugated with a golden MNP.  Then the
renormalized resonance frequency {\it increases} with the excitation
intensity, ranging from $\omega_0 - G_\mathrm{R}$ ($Z = -1$ in the ground
state) to $\omega_0 + G_\mathrm{R}$ ($Z = 1$ in the excited state).  In this
case the response of the system depends on the relative position of the
excitation frequency $\omega$ with respect to the renormalized SQD
transition frequency $\omega_0 - G_\mathrm{R}$.  Thus, if $\omega < \omega_0
- G_\mathrm{R}$ or equivalently $\Delta > G_\mathrm{R}$ (see
Fig.~\ref{Fig2}, left plot) then the excitation is driving the SQD out of
resonance, so that the SQD is becoming less absorptive.  Contrary to that,
if $\omega > \omega_0 - G_\mathrm{R}$ or $\Delta < G_\mathrm{R}$ (see
Fig.~\ref{Fig2}, right plot) the SQD is being driven into the
self-sustaining resonance by the incoming field.  That is the case of the
positive loopback.  In this case, apart from the usual linear ``weak field''
solution, the second kind of a stable state can turn up, which results from
the above-mentioned positive feedback mechanism.  We show below that the
latter has a threshold character, giving rise to bistability and hysteresis
of the system response characteristics.

\section{Steady-state analysis}
\label{Steady-state analysis}

First, we analyze Eq.~(\ref{dotZ}) and Eq.~(\ref{dotR}) under steady-state
conditions ($\dot{Z} = \dot{R} = 0$) to obtain stationary states of the
system.  The corresponding solutions read:
\begin{subequations}
\begin{equation}
\label{steady state Z}
    \frac{|\widetilde\Omega_0|^2}{\gamma\,\Gamma}=
    -\frac{Z+1}{Z}\> \frac{|(\Gamma - G_\mathrm{I}Z) + i\,(\Delta + G_\mathrm{R}Z)|^2} {\Gamma^2}
\end{equation}
\begin{equation}
\label{steady state R}
    R = \frac{Z\;\widetilde\Omega_0}
	{(\Gamma - G_\mathrm{I}\,Z)+i\,(\Delta + G_\mathrm{R}\,Z)}\ .
\end{equation}
\end{subequations}
Equation~(\ref{steady state Z}) is of the third order in $Z$ and therefore
may have three real solutions, depending on values of $\Delta$, $\Gamma$,
$G_\mathrm{R}$, and $G_\mathrm{I}$.  The same applies to the SQD dipole
moment amplitude $R$.

Hereafter, we consider a CdSe SQD in the vicinity of an Au MNP and use the
following set of parameters: the transition energy $\hbar\omega_0 = 2.36$ eV
(which corresponds to the optical transition in a 3.3 nm SQD), the SQD
dielectric constant $\varepsilon_s = 6.2$, the SQD transition dipole moment
$\mu = 0.65$ e$\cdot$nm,~\cite{Zhang06} the MNP radius $a=10$ nm, the
SQD-MNP center-to-center distance $d = 17$ nm, the host dielectric constant
$\varepsilon_b = 1$, and the relaxation constants $\gamma$ and $\Gamma$
defined through $1/\gamma=0.8$ ns and $1/\Gamma=0.3$ ns.~\cite{Artuso08} To
calculate the polarizability $\gamma(\omega)$ of the MNP, we used the
tabulated data for the permittivity of gold from
Ref.~\onlinecite{Johnson72}.  For these parameters $G = G_\mathrm{R} +
i\,G_\mathrm{I} = (25.4 + 10.6\,i)\,\Gamma$.  Note that the frequency domain
of our interest is a narrow region in the vicinity of the SQD resonance,
with the width of about several units of $\Gamma$ (see below), which is much
smaller that the width of the MNP plasmonic resonance.  We therefore
neglected the frequency dependence of the MNP polarizability when
calculating the feedback parameter $G$ and used
$\gamma(\omega)\approx\gamma(\omega_0)$.

\begin{figure}[ht]
\begin{center}
         \includegraphics[width=0.49\columnwidth]{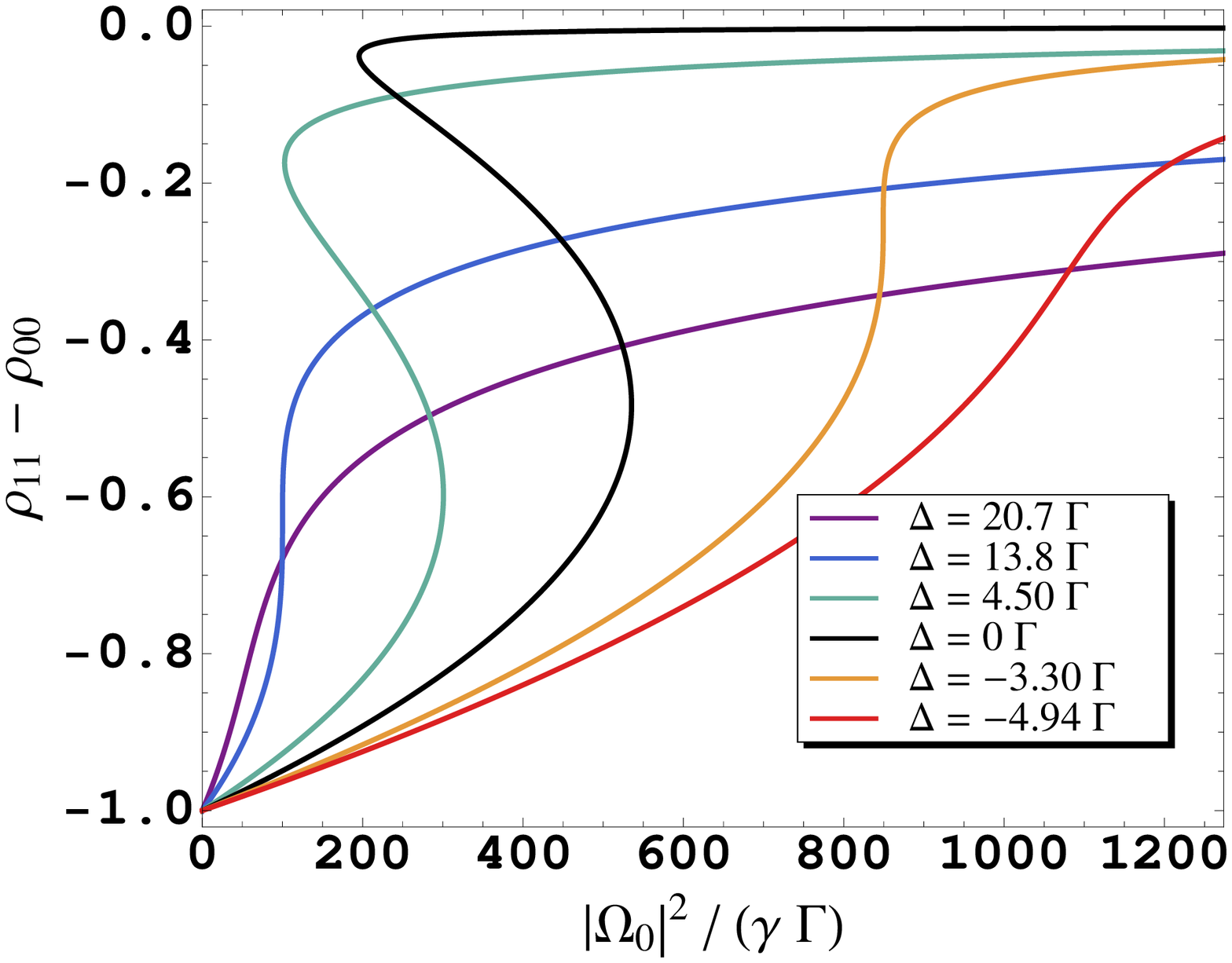}
%\end{center}
%\begin{center}
         \includegraphics[width=0.49\columnwidth]{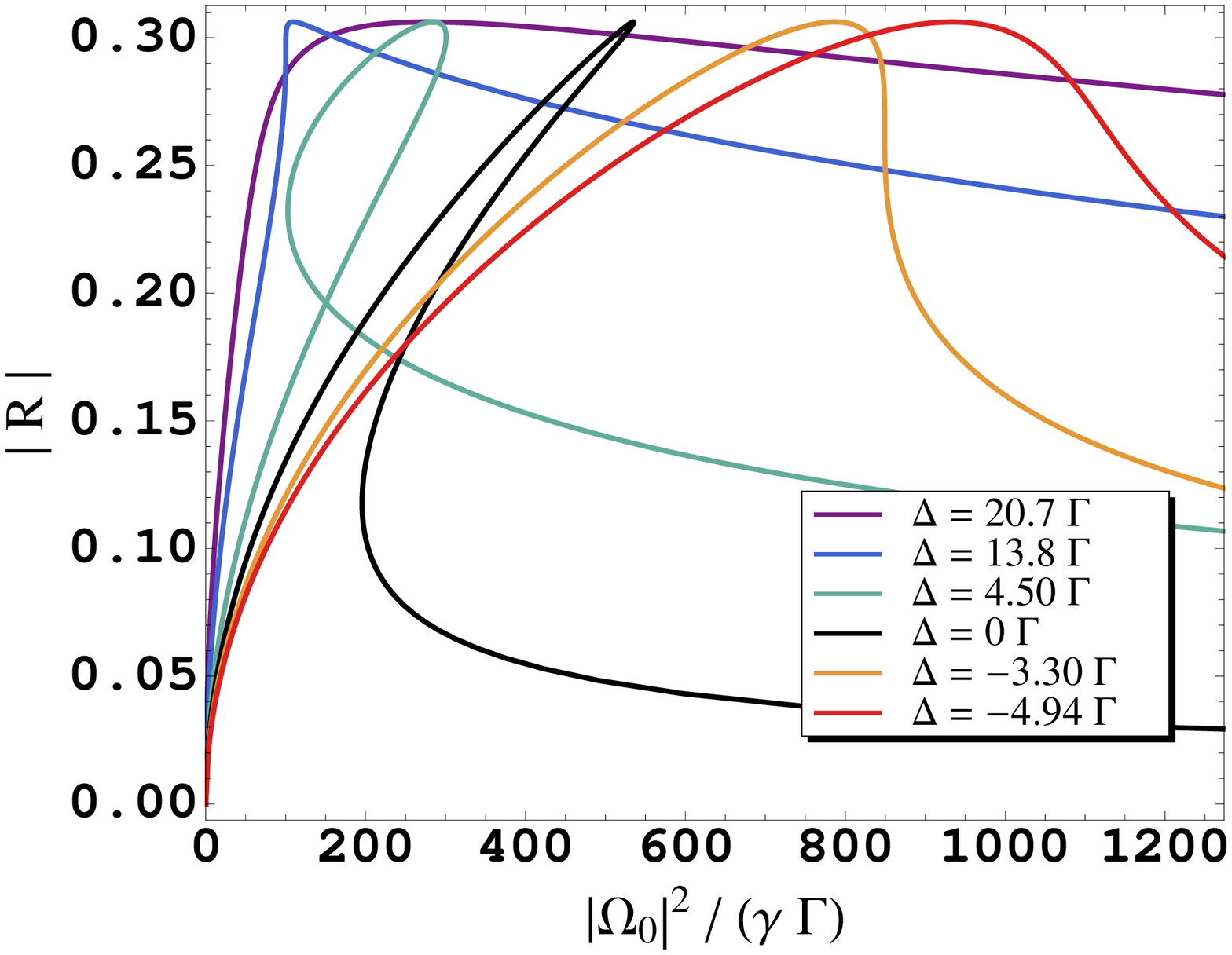}
\end{center}
\begin{center}
         \includegraphics[width=0.49\columnwidth]{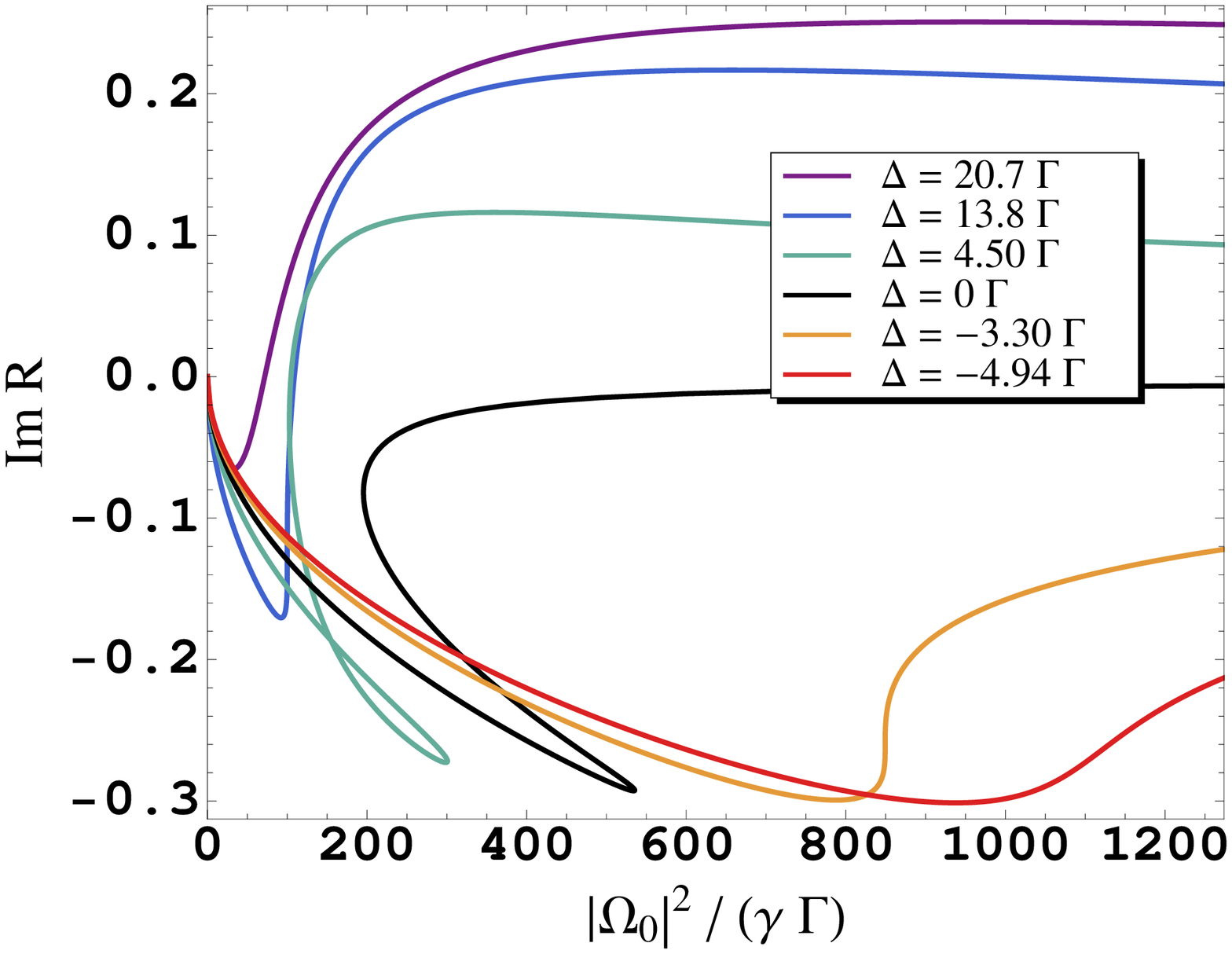}
%\end{center}
%\begin{center}
         \includegraphics[width=0.49\columnwidth]{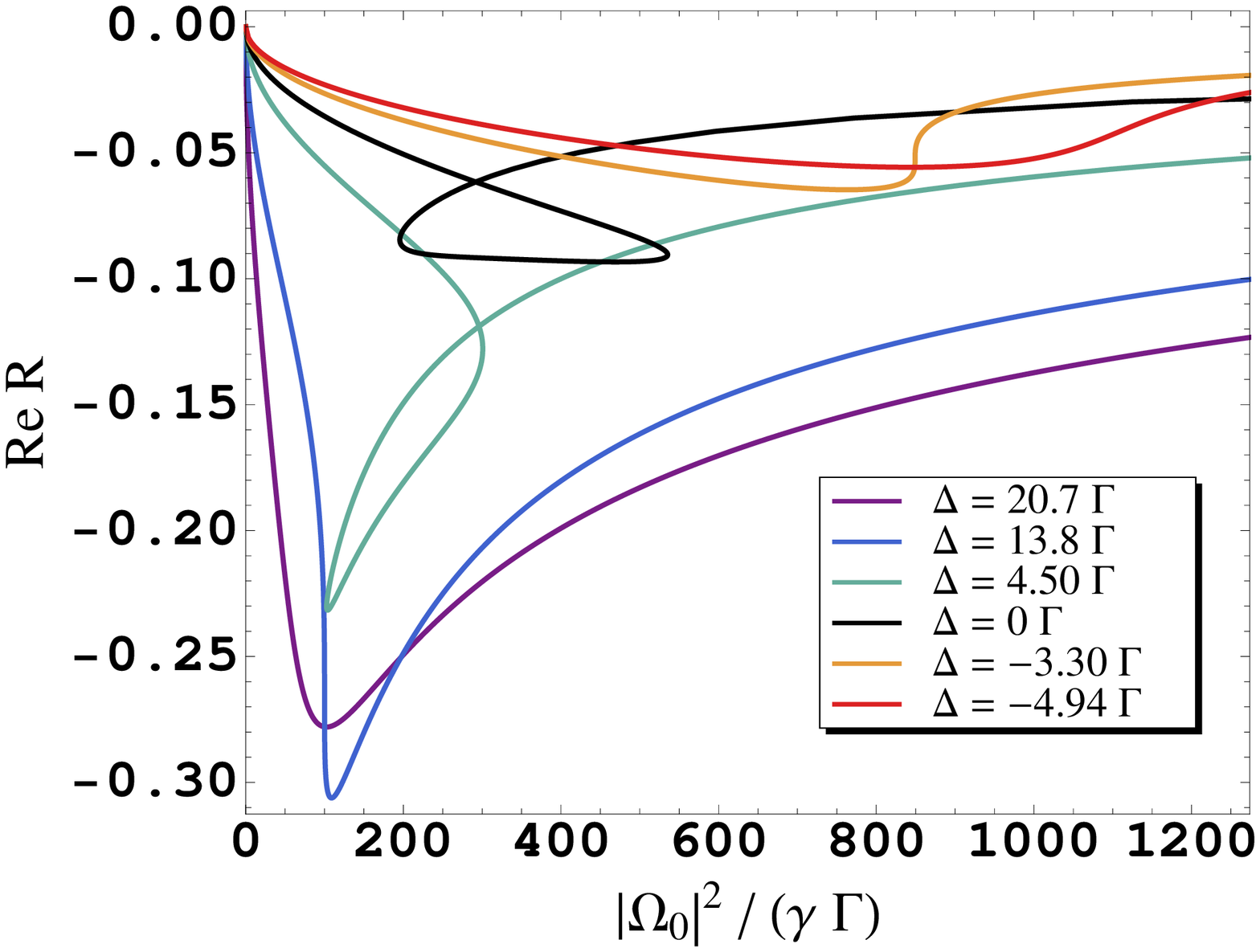}
\end{center}
    \caption{Stationary solutions to Eq.~(\ref{steady state Z}) and
    Eq.~(\ref{steady state R}) for a hybrid nano-molecule comprising a 3.3
    nm CdSe SQD and 20 nm Au sphere separated by the center-to-center
    distance of 17 nm (other parameters are described in the text).  The
    solutions are calculated for different detuning $\Delta$ as functions of
    the normalized external field intensity $\Omega_0/\sqrt{\gamma\Gamma}$. 
    The upper row shows the population difference $Z = \rho_{11} -
    \rho_{00}$ (left) and the SQD dipole moment amplitude $|R|$ (right). 
    The lower left panel displays the absorptive part of the dipole
    amplitude, Im($R$), while the right one --- its dispersive part,
    Re($R$).
} \label{PlotZonOmega}
\end{figure}

Figure~\ref{PlotZonOmega} shows the solution of Eq.~(\ref{steady state Z})
and Eq.~(\ref{steady state R}) for the set of parameters specified above and
different detunings $\Delta$.  As is seen from the plots, within a window of
$-3.3\,\Gamma\leq\Delta\leq 13.8\,\Gamma$, the field dependence of $Z$ and
$R$, have three allowed values for a given intensity
$|\Omega_0|^2/(\gamma\Gamma)$.  The upper limit of the window, $\Delta =
13.8\,\Gamma$, corresponds to the excitation frequency $\omega = \omega_0 -
13.8\,\Gamma$ which lies above the renormalized SQD resonance frequency
$\omega_0 - G_\mathrm{R} = \omega_0 - 25.4\,\Gamma$ (the positive loopback
case).  The lower limit of the window, $\Delta = -3.3\,\Gamma$, is negative
and the corresponding frequency lies above the bare resonance. 
Nevertheless, the SQD can still be driven into the self-sustaining resonance
by the external field.  At larger $\omega$ (larger negative $\Delta$), the
resonance between the excitation and the SQD can not be attained because it
requires a significant positive population difference $Z$ that is
unreachable under stationary conditions: due to the saturation effect, the
upper limit for the population difference is $Z=0$ in the steady state. 
Because of that, the bistability effect disappears for large negative
detunings.

Optical bistability can therefore be observed within a window of detunings
in the vicinity of the SQD transition frequency.  The width of the window is
typically on the order of several units/few dozens of $\Gamma$.  In small
SQDs the dipole dephasing time is usually strongly temperature dependent and
can change from ns for low temperatures to less than ps for room
temperature.~\cite{Alivisatos01} The bistability exists only if the feedback
parameter $G$ exceeds some threshold value.  If $G_\mathrm{I} = 0$ the
condition for bistability to occur is $G_\mathrm{R} \ge
4\Gamma$~\cite{Friedberg89} while if $G_\mathrm{R} = 0$ the condition is
$G_\mathrm{I} \ge 8\Gamma$.~\cite{Basharov88} The value of $G$ is determined
by the geometry and material properties and can not be increased
arbitrarily.  Therefore, these criteria can be used to choose suitable
materials and system configuration.

We note that only the population difference $Z$ manifests the standard
S-shape curves while all the quantities related to the SQD dipole amplitude
$R$ exhibit more exotic coiled curves.  As we show below, the latter leads
to completely different types of hysteresis loops for these quantities.

\begin{figure}[ht]
\begin{center}
\includegraphics[width=0.49\columnwidth]{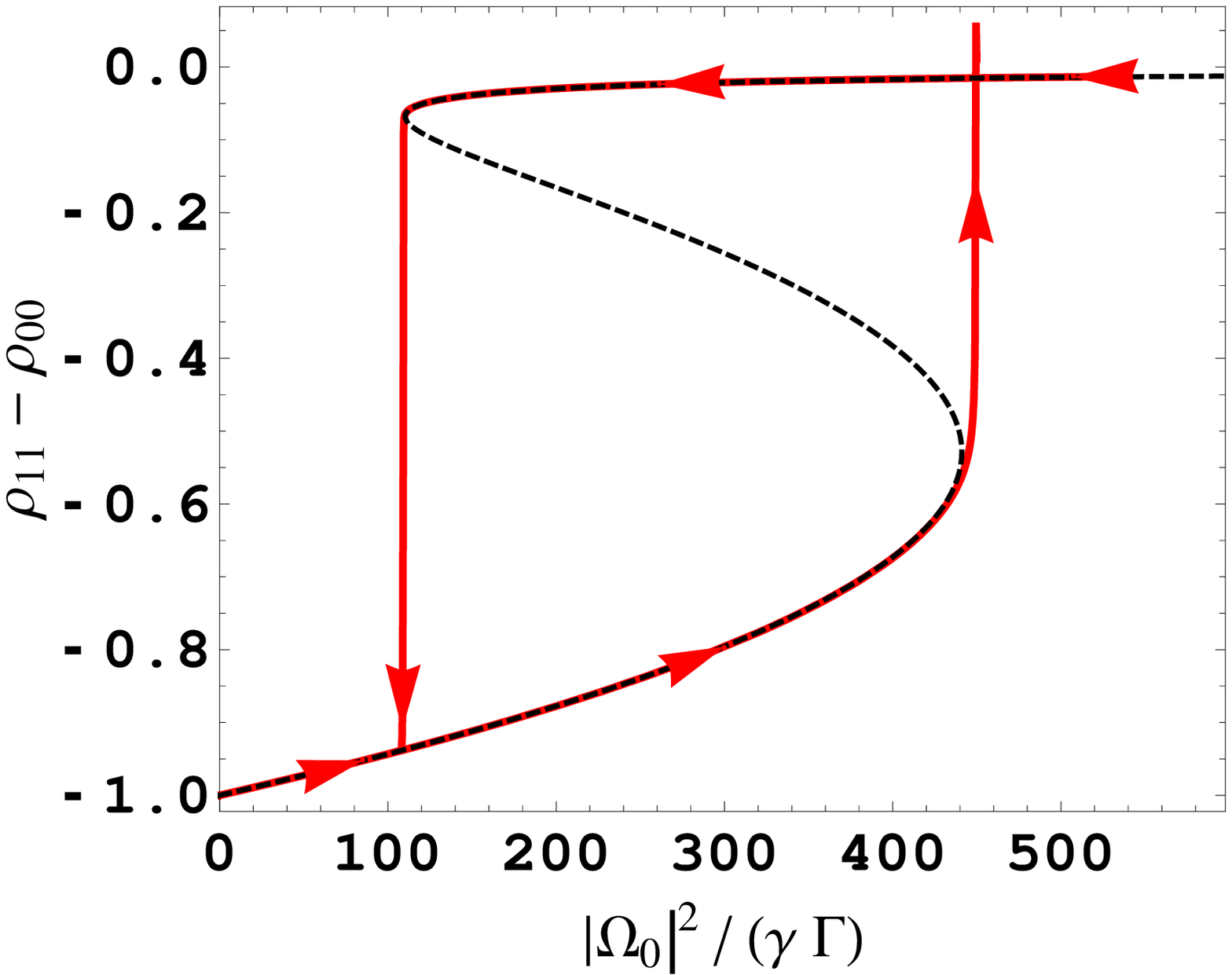}
%\end{center}
%\begin{center}
\includegraphics[width=0.49\columnwidth]{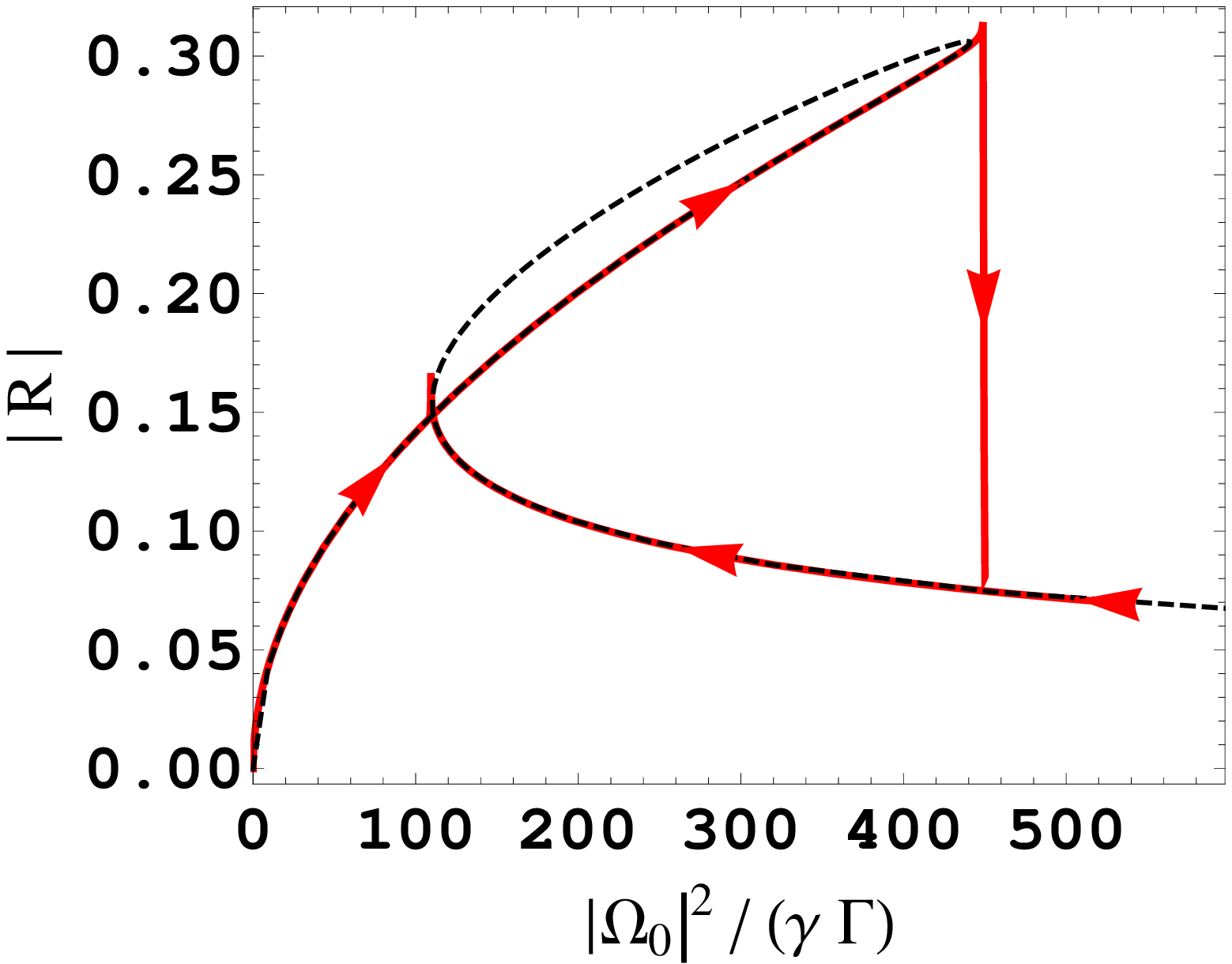}
\end{center}
\begin{center}
\includegraphics[width=0.49\columnwidth]{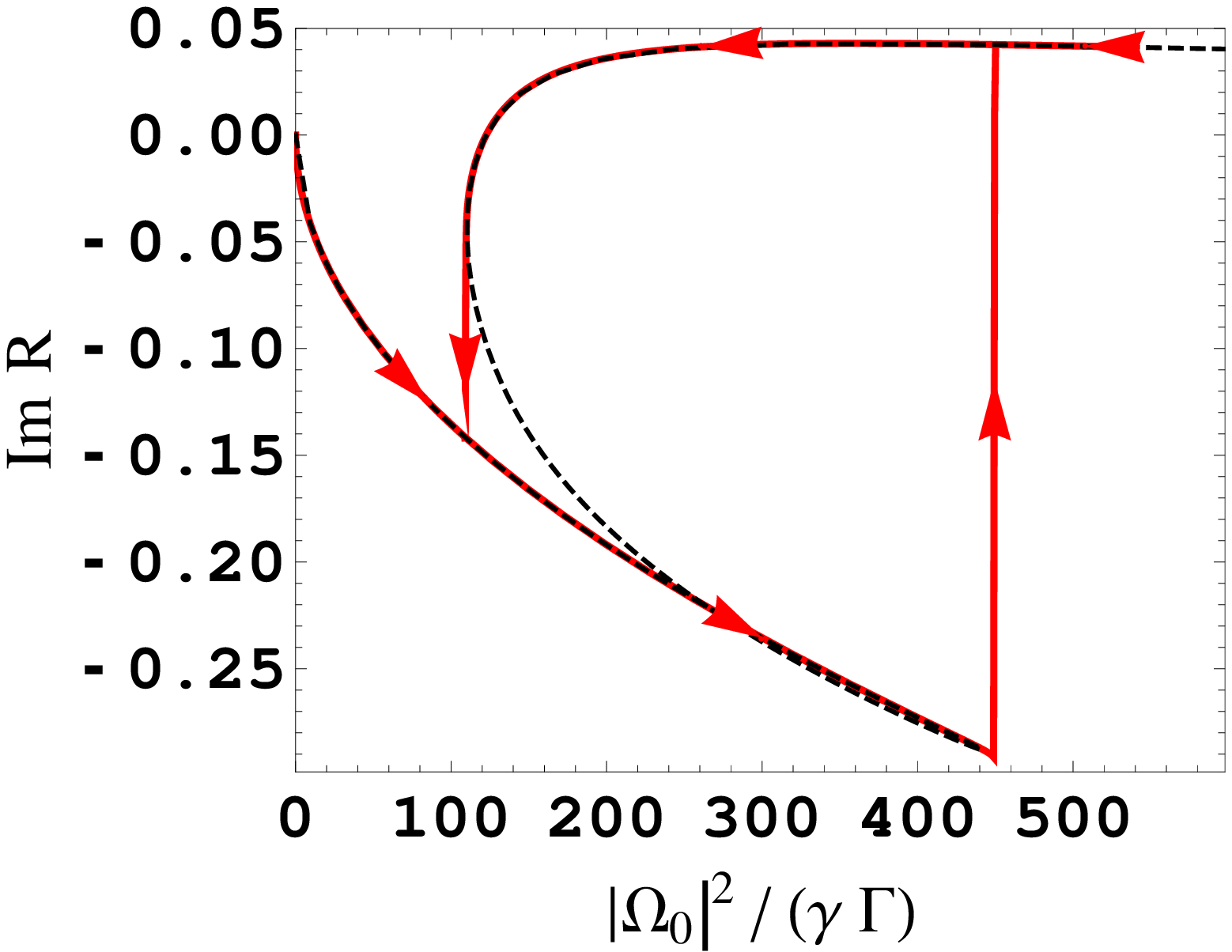}
%\end{center}
%\begin{center}
\includegraphics[width=0.49\columnwidth]{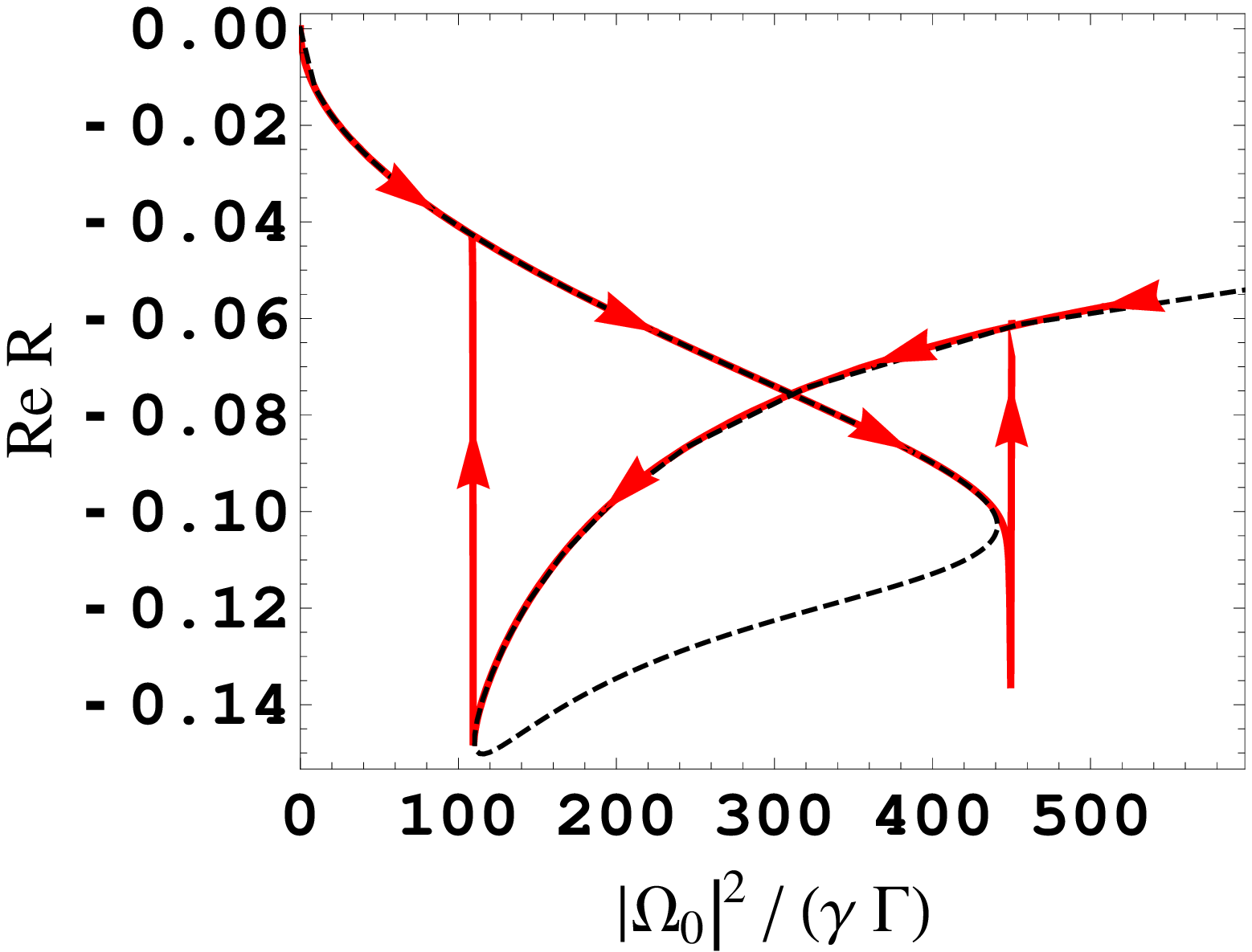}
\end{center}
\caption{Optical hysteresis loops of the population difference $Z =
\rho_{11} - \rho_{00}$ and the SQD dipole moment amplitude $R$ obtained by
solving Eqs.~(\ref{dotZ}) and (\ref{dotR}) with the external field intensity
${|\Omega_0|^2/\sqrt{\gamma\Gamma}}$ being adiabatically swept back and
forth across the bistability region (the sweeping speed of the renormalized
Rabi frequency $\widetilde\Omega_0/\sqrt{\gamma\Gamma}$ is $2.1\times
10^{-3}\Gamma$) The detuning from the resonance is $\Delta = 1.5\Gamma$. 
All other parameters are as in Fig.~\ref{PlotZonOmega}.}
\label{Hysteresis}
\end{figure}

\section{Optical hysteresis}
\label{Optical hysteresis}

We performed time-domain calculations with the external field intensity
$|\Omega_0|^2/(\gamma\Gamma)$ being adiabatically swept back and forth
across the bistability region (the sweeping speeds are given in figure
captions), monitoring the evolution of the system to figure out which steady
state branches are stable.  The results are shown in Fig.~\ref{Hysteresis}. 
For the population difference $Z$ and the absorptive part of the SQD dipole
moment Im$(R)$ we find the standard behavior.  Upon increasing the applied
intensity, the system follows the lower (stable) branch until the intensity
reaches the critical value at which the system switches to the upper branch
(which is also stable).  Upon sweeping the intensity back, the system stays
on the upper branch and then switches back down to the lower one at the
other critical intensity, completing the hysteresis loop.  The intermediate
branch can not be revealed by the adiabatic sweeping of the field because it
is unstable which can also be checked by the standard stability analysis.

Both the $|R|$ and the Im($R$) behave in a very different manner,
manifesting hysteresis loops with kinks.  In the case of the $|R|$ the upper
branch is unstable and the hysteresis loop is triangular, while the Im($R$)
with its unstable lower branch has even more complicated bow-tie hysteresis
curve.  To the best of our knowledge no such optical hysteresis loops have
neither been predicted nor observed so far.

These unusual properties are also expected to manifest themselves in the MNP
dipole moment because it depends on that of the SQD [see Eq.~(\ref{P_MNP})] and
can therefore be switched abruptly as well.

\begin{figure}[ht!]
\begin{center}
\includegraphics[width=0.49\columnwidth]{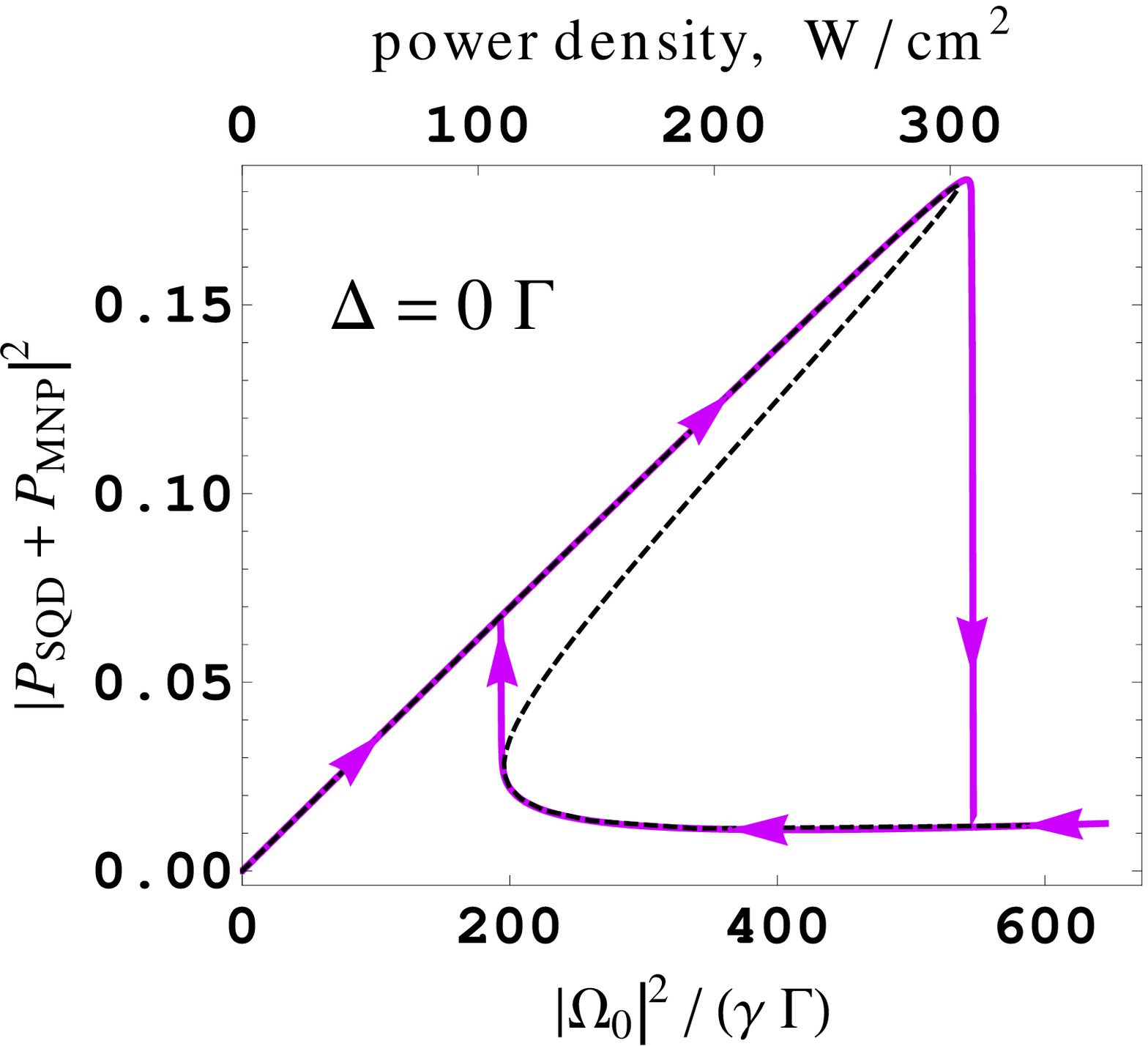}
\includegraphics[width=0.49\columnwidth]{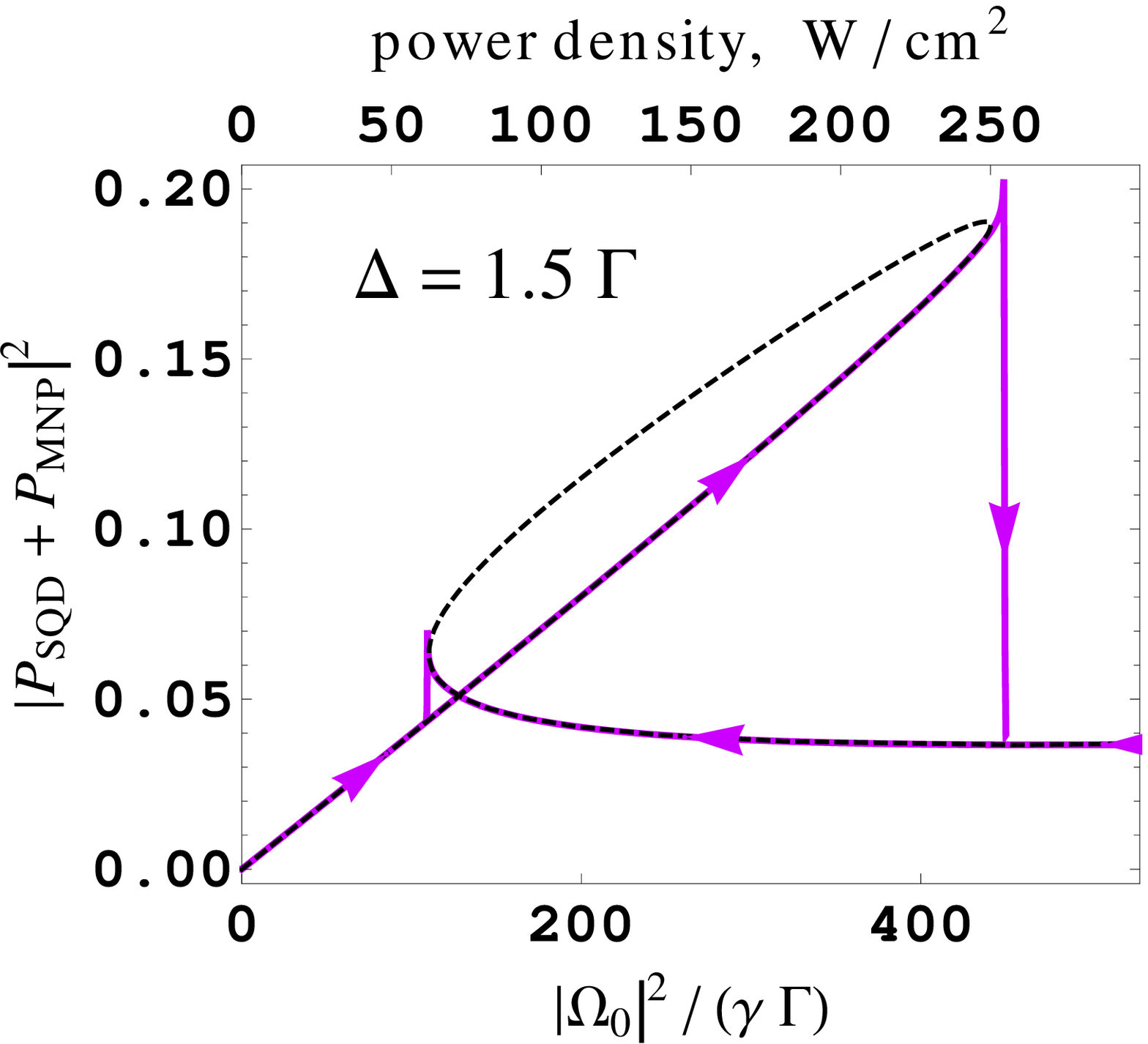}
\vskip 10pt
\includegraphics[width=0.49\columnwidth]{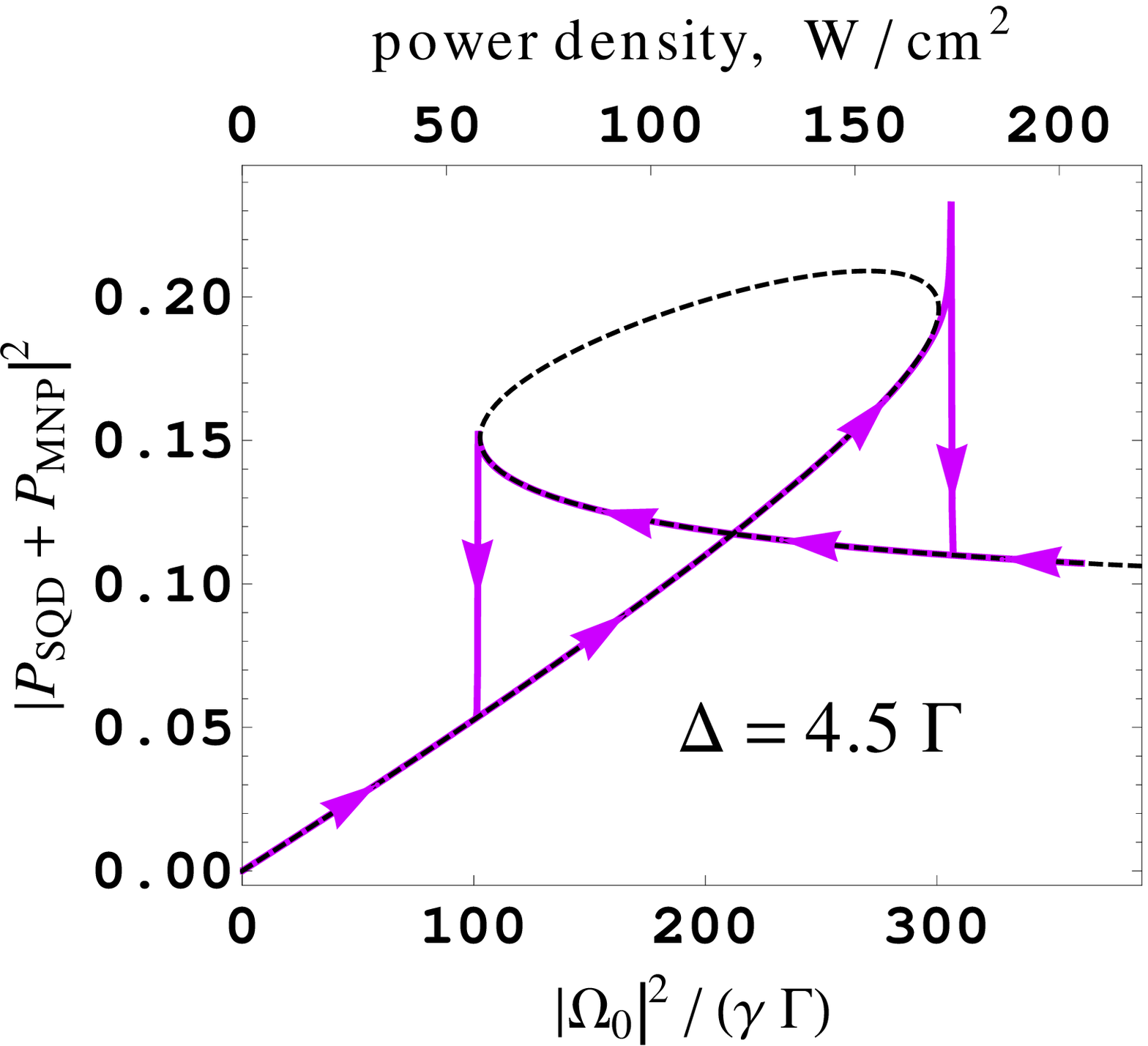}
\includegraphics[width=0.49\columnwidth]{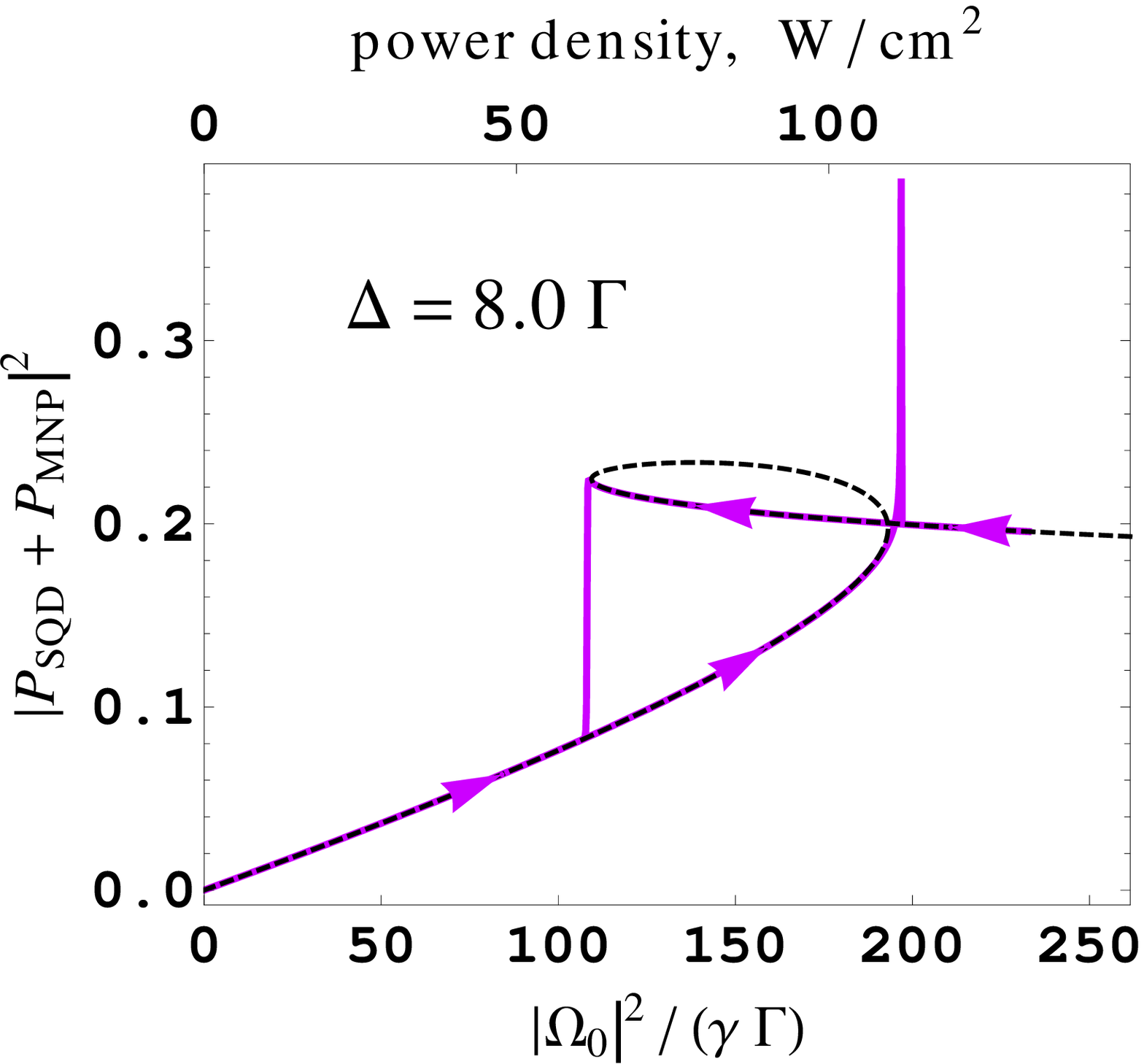}
\end{center}
    \caption{The Rayleigh scattering intensity $|P_\mathrm{SQD} +
    P_\mathrm{MNP}|^2$ in units of $|\mu|^2$ for different values of
    $\Delta$ (indicated in the figures) as a function of excitation
    intensity. The corresponding power density is shown in the upper axis.
    Sweeping speeds of the renormalized Rabi frequency
    $\widetilde\Omega_0/\sqrt{\gamma\Gamma}$ in units of $\Gamma$
    is on the order of $10^{-3}$ for all values of $\Delta$.
    %are
    %$2.4\times10^{-3}$ ($\Delta=0$), $2.2\times10^{-3}$
    %($\Delta=1.5\Gamma$), $1.8\times10^{-3}$ ($\Delta=4.5\Gamma$),
    %$1.6\times10^{-3}$ ($\Delta=10\Gamma$).
    All other parameters are as in Fig.~\ref{PlotZonOmega}.}
\label{Rayleigh scattering}
\end{figure}

\section{Rayleigh scattering and optical storage}
\label{Rayleigh scattering and optical storage}

In experiment, the intensity of the Rayleigh scattering can be measured. 
The amplitude of this characteristic is known to be proportional to the
squared absolute value of the total system dipole moment $|P_\mathrm{SQD} +
P_\mathrm{MNP}|^2$ and is therefore expected to manifest bistability as
well.  We plot the amplitude of the Rayleigh scattering in
Fig.~\ref{Rayleigh scattering} for different values of the detuning
$\Delta$.  The figure shows that various types of hysteresis curves can be
observed: the standard loop (for $\Delta=0$) as well as more exotic
triangular loops ($\Delta=1.5\,\Gamma$ and $\Delta=8\,\Gamma$) and the
bow-tie one ($\Delta=4.5\,\Gamma$).  The two latter types of hysteresis
curves are characteristic for the SQD dipole moment (see
Fig.~\ref{Hysteresis}), which suggests that the major contribution to the
scattering is coming from the SQD.  To confirm the latter we calculated the
relative contribution of the SQD dipole moment into the total scattering
amplitude, $|P_\mathrm{SQD}|^2/|P_\mathrm{SQD} + P_\mathrm{MNP}|^2$ which
turned up to be on the order of unity across the whole hysteresis region. 
Within this region the contribution of the MNP to the scattered field is
typically an order of magnitude less than that of the SQD for the chosen set
of parameters.

It is to be noticed that the ratio of the scattering intensity in the two
stable states can be as high as about $25$ for $\Delta=0$.  Such high
contrast suggests that the system can be used as an optical memory cell:
lower intensity can represent a logical zero while the higher -- a logical
one.  The state of the cell can be switched by sweeping the excitation power
across the bistability region.  Another possibility to switch the cell is to
maintain the power constant while changing the incident field polarization. 
The latter mechanism is characteristic for this particular system due to its
axial symmetry.  We point out that such memory is a volatile one as it
requires constant pumping.

\section{Summary}
\label{Summary}

We investigated theoretically the optical response of a hybrid ``artificial
molecule'' comprised of a closely spaced spherical semiconductor quantum dot
(modeled as a two-level system) and a metal nanosphere (considered
classically), which are coupled by the dipole-dipole interaction.  The
interaction results in self-action of the SQD via the MNP, leading to the
population dependence of the SQD transition frequency and relaxation
constant of the SQD dipole moment.  This provides a feedback mechanism
resulting in several fascinating effects.  Thus, we found that the system
can manifest bistability and optical hysteresis.  In particular, the total
dipole moment of the system can be switched between its two stable states by
the incoming field.  The latter suggests such possible applications as
optical memory cells and all-optical switches at nanoscale in the visible.

Because the SQD-MNP dipole-dipole interaction depends orientation of the
dipole moments of the two particles, the switching can be achieved not only
by the traditional control by the incident amplitude, but also by changing
the polarization of the incoming field with respect to the system axis.  Our
calculations performed for typical system parameters, such as those of a
CdSe or CdSe/ZnSe quantum dot and an Au nanoparticle complexes, predict the
optical bistability of a SQD-MNP artificial molecule.  The Rayleigh
scattering and modern methods of single molecule/particle
spectroscopy~\cite{Alivisatos01,Anderson10,Tcherniak10,Slaughter10} could
probably be used to discover the predicted effects experimentally.

To conclude, we considered the simplest diatomic hybrid artificial
nano-molecule.  We expect, however, that more complicated clusters (such as
an SQD surrounded by several MNPs, as considered in
Ref.~\onlinecite{Govorov06}) can also exhibit these effects because in such
systems nanoparticles are just playing the role of a ``resonator'' and
provide feedback to the nonlinear two-level system.  Anisotropy of
nanoparticles can also easily be accounted for by using an appropriate
tensor instead of the scalar polarizability.  Finally, we note that a very
interesting aspect of this kind of systems is the direction of the total
induced dipole moment, which can also be bistable.
%
%However, the study of this problem goes beyond the scope of this paper and
%will be published elsewhere.

\section{Acknowledgments}

A. V. M. acknowledges support from projects MOSAICO (FIS2006-01485),
BSCH-UCM (PR58/08), the Ram\'on y Cajal program (Ministerio de Ciencia e
Innovaci{\'o}n de Espa{\~n}a) and is grateful to the University of Groningen
for hospitality.

\end{document}